\documentclass[showpacs, prd, draft, twocolumn, nofootinbib, preprintnumbers,
floatfix]{revtex4}

\usepackage{amsmath}
\usepackage{amssymb}
\usepackage{color}
\usepackage{verbatim}

\newcommand{\dif}{\mathrm{d}}
\newcommand{\mc}{\mathcal{C}}
\newcommand{\fc}{\mathfrak{C}}
\newcommand{\md}{\mathcal{D}}

\newcommand{\ba}{\bar{A}}

\begin{document}

\title{Vector models of gravitational Lorentz symmetry breaking}

\author{Michael D.\ Seifert}

\affiliation{Dept.\ of Physics, Indiana University, 727 E.\
$\text{3}^\text{rd}$ St., Bloomington, IN, 47405}

\email{mdseifer@indiana.edu}

\preprint{IUHET 526, March 2009}

\pacs{03.50.-z, 04.25.Nx, 04.50.Kd}


\begin{abstract}
  Spontaneous Lorentz symmetry breaking can occur when the dynamics of
  a tensor field cause it to take on a non-zero expectation value
  \textit{in vacuo}, thereby providing one or more ``preferred
  directions'' in spacetime.  Couplings between such fields and
  spacetime curvature will then affect the dynamics of the metric,
  leading to interesting gravitational effects.  Bailey \&
  Kosteleck\'{y} \cite{BK} developed a post-Newtonian formalism that,
  under certain conditions concerning the field's couplings and
  stress-energy, allows for the analysis of gravitational effects in
  the presence of Lorentz symmetry breaking.  We perform a systematic
  survey of vector models of spontaneous Lorentz symmetry breaking.
  We find that a two-parameter class of vector models, those with
  kinetic terms we call ``pseudo-Maxwell,'' can be successfully
  analyzed under the Bailey-Kosteleck\'{y} formalism, and that one of
  these two ``dimensions'' in parameter space has not yet been
  explored as a possible mechanism of spontaneous Lorentz symmetry
  breaking.
\end{abstract}


\maketitle

\section{Introduction}

It is widely believed that classical general relativity, as formulated
by Einstein, is a particular limit of some underlying theory of
quantum gravity.  However, at energy scales that are now accessible,
it is expected (from our knowledge of effective field theory) that any
fundamentally non-classical effects would be suppressed by at least a
factor of the ratio of our experimental energy scale to the Planck
scale; even for today's most powerful particle colliders, this ratio
still gives a suppression factor of $10^{-16}$.  With no foreseeable
way to bridge this sixteen-order-of-magnitude gap in energy, we are
forced to aim for sensitivity rather than power when searching for
quantum-gravitational effects.

One particularly interesting avenue for this search is the possibility
of quantum-suppressed Lorentz violation.  In such a scenario, the
underlying theory would include a tensor field (or fields) which
spontaneously takes on a non-zero expectation value.  Such a field
would, in essence, provide a ``preferred'' direction or directions in
spacetime.\footnote{Such a field is often said to be
  ``Lorentz-violating''.  This description plays somewhat fast and
  loose with usual notions from the rest of physics; the field does,
  after all, transform as a proper tensor field under local Lorentz
  transformations.  A more accurate way to describe such a field would
  be to say that it ``spontaneously breaks Lorentz symmetry'', but
  such phrasing is rather more awkward.  In the interests of
  readability and consistency with other papers in the literature, we
  will use ``Lorentz-violating'' in this sense as well.}  The
background value of this field could then couple weakly to
conventional matter fields \cite{Kosgrav}; thus, the effects of such a
tensor field could in principle be seen via careful observation of the
behaviour of conventional particles and fields.

A particularly interesting venue in which to search for possible
violations of Lorentz invariance is the gravitational sector.
Interactions between a dynamical metric and a tensor field with a
non-zero expectation value have been postulated as a possible method
of modifying cosmology
\cite{aethmicro,sigaeth,instaeth,LVcosmo,JanSoda}, as a mechanism for
modifying Newtonian gravity to solve the dark-matter problem
\cite{TeVeS, TeVesvec}, or simply in their own right as modifications
of conventional gravity \cite{KosSam,einaeth,cardgrav}.  Such
modifications of gravity will, in general, cause modifications to the
weak-field limit of gravity.  The linearized effects of a direct
coupling between Lorentz-violating fields and the Riemann tensor were
analyzed in some detail by Bailey and Kosteleck\'y \cite{BK}.  By
making certain assumptions about the properties of the equations of
motion, they were able to obtain an effective linearized gravitational
equation of the form
\begin{equation}
  \label{effgrav}
  \delta G_{ab} + \mathfrak{T}_{ab} {}^{cdef} \delta R_{cdef} = 8 \pi
  G \delta T_{ab} 
\end{equation}
where $\delta G_{ab}$ and $\delta R_{abcd}$ are (respectively) the
Einstein and Riemann tensors linearized about a flat background,
$\delta T_{ab}$ is the stress-energy of conventional matter, and
$\mathfrak{T}_{ab} {}^{cdef}$ is a ``small'' tensor (in a sense we
will make explicit below) depending in a particular way on the
background values of the Lorentz-violating tensors.  Using this
effective equation, they then performed a thorough post-Newtonian
analysis of such theories, examining the effects of Lorentz-violating
fields on phenomena including satellite orbits, interferometric
gravimetry, torsion-balance experiments, and
frame-dragging.\footnote{It is important to note that although the
  Bailey-Kosteleck\'y formalism can be applied to the analysis of
  post-Newtonian gravity, the theories to which this formalism can be
  applied are in general not the same as those to which Will's
  familiar Parametrized Post-Newtonian (PPN) formalism \cite{Willbook,
    Willrev} can be applied.  The connections and distinctions between
  these two formalisms are explored in Section III C of Bailey and
  Kosteleck\'y's original paper \cite{BK}. }

While this formalism is highly valuable for the analysis of the
interface between gravity and Lorentz violation, its range of
applicability is not immediately clear.  To obtain the effective
gravitational equation \eqref{effgrav}, it was necessary for Bailey
and Kosteleck\'y to place certain conditions on the equations of
motion, rather than on the action from which they were derived.  As
action principles tend to be conceptually simpler than the equations
of motion derived from them, it would be quite helpful to know whether
a given action which includes spontaneous Lorentz symmetry breaking is
analyzable in the Bailey-Kosteleck\'y formalism.  Should this be the
case, the physical predictions of their paper \cite{BK} would be
directly applicable to any such model.

This question is the focus of the present work.  We will restrict our
attention to the simplest type of tensor field which can spontaneously
break Lorentz symmetry, namely vector fields $A^a$.  In Section
\ref{formalsec}, we describe the properties of the theories we will be
concerned with, and we review the conditions required for successful
use of the Bailey-Kosteleck\'y formalism.  Section \ref{applsec} is
dedicated to the application of these conditions to the vector actions
under consideration; we will see that the class of vector theories for
which the Bailey-Kosteleck\'y formalism can successfully be used is
not large, but that there do exist previously unconsidered models
which can be analyzed in this framework.  Finally, we discuss these
results in Section \ref{discsec}.

We use the sign conventions of Wald \cite{Wald} throughout, and units
in which $c = 1$.

\section{Equations of Motion and Formalism}
\label{formalsec}

\subsection{Actions for Lorentz-breaking vector fields}

Bailey and Kosteleck\'{y}'s analysis of gravitational Lorentz violation
\cite{BK} begins by assuming an action of the form
\begin{equation}
  \label{BKaction}
  S = \int \dif^4 x \sqrt{-g} \left( \mathcal{L}_{EH} +
  \mathcal{L}_{LV} + \mathcal{L}' \right).
\end{equation}
$\mathcal{L}_{EH}$ here is the usual Einstein-Hilbert action,
\begin{equation}
  \mathcal{L}_{EH} = R - 2 \Lambda.
\end{equation}
We will assume throughout that $\Lambda = 0$.  The second term,
$\mathcal{L}_{LV}$, contains the non-trivial couplings of the
Lorentz-violating fields to the metric:
\begin{equation}
  \mathcal{L}_{LV} = - u R + s^{ab}(R^T)_{ab} + t^{abcd} C_{abcd}.
\end{equation}
Here, $R$ is the Ricci scalar, $(R^T)_{ab} \equiv R_{ab} - \frac{1}{4}
g_{ab} R$ is the trace-free Ricci tensor, and $C_{abcd}$ is the Weyl
tensor.  The tensors $u$, $s^{ab}$, and $t^{abcd}$ may be fundamental
fields or (as will be the case in our analysis) composites of other
fields present in the theory.  The final term, $\mathcal{L}'$,
contains the terms determining the dynamics of the fundamental
Lorentz-violating fields, as well as the action for conventional
matter.

In the case of a single vector field being responsible for Lorentz
symmetry breaking, we can be more specific in the form of the
Lagrangian.  Denoting the Lorentz-breaking vector field by $A^a$, the
most general Lorentz-violation coupling terms will be of the form
\begin{equation}
  \label{genericLV}
  \mathcal{L}_{LV} = \xi( -f_u(A^2) R + f_s(A^2) A^a A^b R_{ab} ),
\end{equation}
where $A^2 = A^a A_a$, $f_u$ and $f_s$ are arbitrary functions of
$A^2$, and $\xi$ is a coupling constant.\footnote{Note that $f_s$ is
  associated with the Ricci tensor in our parametrization, while in
  Bailey \& Kosteleck\'y's original paper the tensor $s^{ab}$ is
  associated with the trace-free Ricci tensor.}  (By the symmetries of
the Weyl tensor, any term analogous to $t^{abcd} C_{abcd}$ and
constructed out of $A^a$ and the metric must vanish.)  This term is
best thought of as a ``weak'' coupling term between the vector field
and the curvature;  the ``weakness'' of this coupling will be of
importance in the next subsection.

The dynamics for $A^a$, meanwhile, will be determined by
$\mathcal{L}'$.  We can write the Lagrangian for an arbitrary
second-differential-order vector theory as
\begin{equation}
  \mathcal{L}' = K^{a} {}_b {}^c {}_d \nabla_a A^b \nabla_c A^d -
  V(A^2) + 2 \kappa \mathcal{L}_\text{mat},
\end{equation}
where $\mathcal{L}_\text{mat}$ is the Lagrangian for ``conventional''
matter; $\kappa = 8 \pi G$; $V(A^2)$ is the potential for the vector
field, constructed to have a minimum at a non-zero value of $A^a$; and
$K^a {}_b {}^c {}_d$ is a tensor constructed out of $A^a$ and the
metric.  This tensor can be taken to be symmetric under the
simultaneous exchange of $a \leftrightarrow c$ and $b \leftrightarrow
d$.  The conventional matter action $\mathcal{L}_\text{mat}$ can, in
principle, contain direct couplings to $A^a$. (We will introduce an
explicit parametrization for $K^a {}_b {}^c {}_d$ in the next
subsection.)

We can easily obtain the Euler-Lagrange equations associated with this
action by varying the action with respect to $g_{ab}$ and $A^a$;
there result the equations
\begin{equation}
  \label{fullEineq}
  (\mathcal{E}_g)^{ab} \equiv -G^{ab} + \xi \mathcal{A}^{ab} +
  \xi \mathcal{B}^{ab} + (T_A)^{ab} + \kappa (T_\text{mat})^{ab}= 0
\end{equation}
and
\begin{multline}
  \label{fullveceq}
  (\mathcal{E}_A)_a \equiv 2 \xi ( -f'_u A_a R + f'_s A^b A^c R_{bc} A_a +
  f_s A^b R_{ba}) \\ + \mathcal{M}^b {}_c {}^d {}_e {}_a \nabla_b A^c
  \nabla_d A^e - 2 \nabla_b \left( K^b {}_a {}^c {}_d \nabla_c A^d
  \right) \\ {} - 2 V' A_a + \frac{\delta \mathcal{L}_\text{mat}}{\delta
    A^a}= 0, 
\end{multline}
where 
\begin{multline}
  \mathcal{A}^{ab} \equiv  f_u G^{ab} + f'_u A^a A^b R
  \\ {} + \frac{1}{2}
  f_s g^{ab} A^c A^d R_{cd} + f'_s A^a A^b A^c A^d R_{cd},
\end{multline}
\begin{multline}
  \mathcal{B}^{ab} \equiv (g^{ab} \Box - \nabla^a \nabla^b) f_u -
  \frac{1}{2} g^{ab} \nabla_c \nabla_d (f_s A^c A^d) \\ {} - \frac{1}{2}
  \Box (f_s A^a A^b) + \nabla_c \nabla^{(a} (f_s A^{b)} A^c),
\end{multline}
\begin{multline}
(T_A)^{ab} \equiv \mathcal{M}^c {}_d {}^e {}_f {}^{ab} \nabla_c A^d
  \nabla_e A^f \\ {} + \nabla_e \left( ( K^c {}_d {}^{(a|e|} A^{b)} - K^c
    {}_d {}^{e(a} A^{b)} - K^{(ab)c} {}_d A^e ) \nabla_c A^d \right)\\
  {} -
  \frac{1}{2} g^{ab} V - A^a A^b V',
\end{multline}
\begin{equation}
(T_\text{mat})^{ab} \equiv \frac{1}{2}\frac{1}{\sqrt{-g}} \frac{ \delta
  (\sqrt{-g} \mathcal{L}_\text{mat} )}{\delta g_{ab}},
\end{equation}
\begin{equation}
  \mathcal{M}^c {}_d {}^e {}_f {}^{ab} \equiv \frac{1}{2} g^{ab}
  K^c {}_d {}^e {}_f + \frac{\delta K^c {}_d {}^e {}_f}{\delta g_{ab}},
\end{equation}
and
\begin{equation} 
  \mathcal{M}^b {}_c {}^d {}_e {}_a \equiv \frac{\delta K^b {}_c {}^d
    {}_e}{\delta A^a}.
\end{equation}
(The arguments of the functions $f_u$, $f_s$, and $V$ will be
regularly omitted for brevity hereafter.)

\subsection{Bailey-Kosteleck\'y Formalism}

The basic tack taken by Bailey and Kosteleck\'y in their original
paper \cite{BK} was to start from an action of the form
\eqref{BKaction}, with its associated equations of motion; to
construct the linearized equations of motion about a particular type
of background; and to then impose certain conditions on the background
and the equations of motion such that the linearized equations could
be reduced to a particularly simple form:
\begin{multline}
  \label{sueffgrav}
  \delta G_{ab} = \kappa (\delta T_\text{mat})_{ab} + \bar{u} \delta
  G_{ab} + 
  \eta_{ab} \bar{s}^{cd} \delta R_{cd} \\ - 2 \bar{s}^c {}_{(a} \delta
  R_{b)c} + \frac{1}{2} \bar{s}_{ab} \delta R + \bar{s}^{cd} \delta
  R_{acdb},  
\end{multline}
where $\bar{u}$ and $\bar{s}^{ab}$ are the background values of the
fields $u$ and $s^{ab}$.  We now review and discuss these conditions
as they pertain to the vector theories we are considering.
\begin{enumerate}
\label{condlist}
\item \textit{The background values of the Lorentz-violating fields
    are constant with respect to a background flat spacetime.}  In
  other words, if $\epsilon$ is our linearization parameter, we are
  looking for a family of solutions such that
  \begin{equation}
    \begin{aligned}
      g_{ab} &= \eta_{ab} + \epsilon h_{ab} & A^a = \ba^a + \epsilon
      \tilde{A}^a
    \end{aligned}
  \end{equation}
  with $\ba^a \neq 0$, and, in addition, that
  \begin{equation}
    \nabla_a A^b \sim \mathcal{O}(\epsilon).
  \end{equation}
  We will see below that these requirements constrain the background
  values of $V$, as well as greatly simplifying the equations of
  motion \eqref{fullEineq} and \eqref{fullveceq}.

\item \textit{The dominant Lorentz-violating effects are linear in the
    vacuum values $\bar{u}$, $\bar{s}^{ab}$, and $\bar{t}^{abcd}$.}
  This can be enforced in our case by working only to linear order in
  the coupling constant $\xi$, discarding terms of
  $\mathcal{O}(\xi^2)$ or higher.  Turning this condition around, we
  will also require that in the limit of vanishing $\xi$, the metric
  will obey the Einstein equations; this ensures that our
  ``Lorentz-violating'' perturbed metric will only differ slightly
  from the usual perturbed metric derived from the conventional
  Einstein equations.

\item \textit{The fluctuations $\tilde{u}$, $\tilde{s}^{ab}$, and
    $\tilde{t}^{abcd}$ of the Lorentz-violating fields do not couple
    to the ``conventional matter'' sources.}  This can be ensured by
  demanding that
  \begin{equation}
    \frac{\delta \mathcal{L}_\text{mat}}{\delta A^a} = 0,
  \end{equation}
  thereby eliminating the last term from equation \eqref{fullveceq}
  above.  In essence, this requirement ensures that it is only the
  metric that is directly affected by the dynamical Lorentz breaking.
  ``Conventional'' test particles will still move on geodesics with
  respect to the now-distorted metric, and these distorted paths can
  in principle allow us to indirectly observe the effects of Lorentz
  violation on gravity.  In the remainder of this paper, we will be
  studying ``vacuum solutions'', with all conventional matter sources
  set to zero.

\item \textit{The independently conserved piece of the
    Lorentz-violating stress-energy $(T_A)^{ab}$ vanishes.}  More
  specifically, if we take the divergence of the Einstein equation
  \eqref{fullEineq}, we find that the divergence of $(T_A)^{ab}$ must
  equal the divergence of $\xi(\mathcal{A}^{ab} + \mathcal{B}^{ab})$.
  This relation then allows us to ``reverse-engineer'' the form of
  $(T_A)^{ab}$, up to a piece $\Sigma^{ab}$ whose divergence vanishes.
  This condition is then the statement that $\Sigma^{ab}$ itself
  vanishes.\footnote{Note that this is not strictly speaking necessary
    for the analysis performed by Bailey and Kosteleck\'y to still be
    valid, as noted in the original paper; in fact, it does
    not hold for the bumblebee model \cite{BK}.}


\item \textit{When the Einstein equation \eqref{fullEineq} is
    linearized, any second derivatives of $\tilde{A}^a$ can be
    eliminated from $\mathcal{B}^{ab}$ and $(T_A)^{ab}$ in favour of
    second derivatives of the metric.}  In practise, this elimination
  can only occur via the linearized vector equation of motion.  This
  condition will be our primary focus in Section \ref{applsec}.
\end{enumerate}

As a consequence of the first condition above, the background
(zero-order) equations of motion reduce simply to
\begin{equation}
  \frac{1}{2} \eta^{ab} V(\ba^2) + \ba^a \ba^b
  V'(\ba^2) = 0 
\end{equation}
and
\begin{equation}
  V'(\ba^2) \ba_a = 0,
\end{equation}
which together imply (as would be expected) that $V(\ba^2) =
V'(\ba^2) = 0$.  The linearized Einstein equation of motion then
becomes 
\begin{widetext}
\begin{multline}
  \label{linEineq}
  \delta(\mathcal{E}_g)^{ab} = - \delta G^{ab} + \xi \left( f_u \delta
    G^{ab} - f'_u \ba^a \ba^b \delta R
  + \frac{1}{2} f_s \eta^{ab} \ba^c \ba^d \delta
    R_{cd} + f'_s \ba^a \ba^b \ba^c \ba^d \delta
    R_{cd} \right) \\ 
  + ( \xi \mathcal{Q}_R {}^{abc} {}_d {}^e + \mathcal{Q}_K {}^{abc}
  {}_d {}^e ) \delta( \nabla_e \nabla_c A^d )
  - V'' \ba^a \ba^b ( 2 \ba_c \tilde{A}^c + h_{cd}
  \ba^c \ba^d ),
\end{multline}
where
  \begin{multline}
    \label{QRdef}
    \mathcal{Q}_R {}^{abc} {}_d {}^e \equiv 2 f'_u (\eta^{ab}
    \eta^{ce} - \eta^{e(a} \eta^{b)c}) \ba_d + f_s \left( -
      \eta^{ab} \ba^{(c} \delta^{e)} {}_d - g^{ce} \ba^{(a}
      \delta^{b)} {}_d + \eta^{c(a} \ba^{b)} \delta^e {}_d +
      \eta^{c(a} \delta^{b)} {}_d \ba^e \right) \\ + f'_s
    (-\eta^{ab} \ba^c \ba^e - \ba^a \ba^b \eta^{ce} +
    2 \eta^{c(a} \ba^{b)} \ba^e) \ba_d
    \end{multline}
\end{widetext}
and
\begin{equation}
  \label{QKdef}
  \mathcal{Q}_K {}^{abc} {}_d {}^e \equiv \ba^{(a} K^{b)ec} {}_d -
  \ba^{(a} K^{|e|b)c} {}_d - \ba^e K^{(ab)c} {}_d . 
\end{equation}
The linearized vector equation of motion, meanwhile, becomes
\begin{multline}
  \label{linveceq}
  \frac{1}{2} \delta(\mathcal{E}_A)_a = \xi ( -f'_u \ba_a \delta R
  + f'_s \ba^b \ba^c \delta R_{bc} \ba_a + f_s \ba^b
  \delta R_{ba}) 
  \\ - K^b {}_a {}^c {}_d \delta (\nabla_b \nabla_c A^d) - V'' \ba_a
  ( 2 \ba_b \tilde{A}^b + h_{bc} \ba^b \ba^c ).
\end{multline}
In equations \eqref{linEineq}--\eqref{linveceq}, the arguments of the
functions $f_u$, $f_s$, and $V$, as well as the tensor $K^a {}_b {}^c
{}_d$, are understood to be evaluated at their background values $A^a
\to \ba^a$ and $g_{ab} \to \eta_{ab}$; indices are raised and lowered
by the flat-space metric $\eta_{ab}$.  The quantity $\delta (\nabla_a
\nabla_b A^c) $ is given in terms of flat-space derivatives and the
metric perturbation $h_{ab}$ by
\begin{equation}
  \delta (\nabla_a \nabla_b A^c) = \partial_a \left( \partial_b
    \tilde{A}^c + \left( \partial_{(b} h_{d)}{}^c -
      \frac{1}{2} \partial^c h_{bd} \right) \ba^d \right).\end{equation}
Note that by Condition 1 above, this is an $\mathcal{O}(\epsilon)$
quantity.\footnote{It is also important to note that the flat-space
  derivative operator $\partial_a$ and the covariant derivative
  operator $\nabla_a$ differ only at order $\epsilon$.  In particular,
  this means that the covariant derivative of an
  $\mathcal{O}(\epsilon)$ quantity (such as $\nabla_a A^b$) differs
  from its flat-space coordinate derivative by
  $\mathcal{O}(\epsilon^2)$, which for the purposes of this paper is
  negligible.}  The quantities $\delta R_{ab}$, $\delta G_{ab}$, and
$\delta R = \eta^{ab} \delta R_{ab}$, finally, are the linearized
Ricci tensor, Einstein tensor, and Ricci scalar associated with the
metric perturbation $h_{ab}$.

It will be to our advantage to introduce a concrete parametrization
for the tensor $K^a {}_b {}^c {}_d$.  Any tensor with the proper index
structure constructed out of $A^a$ and the metric will be of the form 
\begin{multline}
  \label{Kdef}
  K^{a} {}_b {}^c {}_d = \fc_1(A^2) g^{ac} g_{bd} + \fc_2(A^2)
  \delta^a {}_b \delta^c {}_d  + \fc_3(A^2)
  \delta^a {}_d \delta^c {}_b \\ + \fc_4(A^2) A^a A^c g_{bd} +
  \frac{1}{2} \fc_5 (A^2) \left( A^a A_d \delta^c {}_b + A^c A^b
    \delta^a {}_d \right)
  \\ + \fc_6(A^2) A_b A_d g^{ac} + \frac{1}{2}
  \fc_7(A^2) \left( A^a A_b \delta^c {}_d + A^c A_d \delta^a {}_b
  \right) \\ + \fc_8 (A^2) A^a A^c A_b A_d,
\end{multline}
(This particular parametrization is due to Zlosnik \textit{et al.}
\cite{TeVesvec}.)  However, due to the geometric identity
\begin{multline}
  \nabla^a \left[ f(A^2) (A_a \nabla_b A^b - A^b \nabla_b A_a) \right]
  \\ =
  f(A^2) \left( (\nabla_a A^a)^2 - \nabla_b A^a \nabla_a A^b - R_{ab}
    A^a A^b \right) \\ + 2 f'(A^2) \left( A^a A^c \nabla_a A_c
    \nabla_b A^b - A^a A^c \nabla_a A^b \nabla_b A_c \right),
\end{multline}
we can always eliminate one of $\fc_2$, $\fc_3$, $\fc_5$, or $\fc_7$
via an integration by parts (thereby changing $f_s$ as well.)
Hereafter we will take $\fc_2$ to vanish.  The arguments of
$\fc_i(A^2)$ will also generally be omitted for brevity.

\subsection{``Pseudo-Maxwell'' kinetic terms}
\label{specialKsec}

Finally, we note two important properties of the vector equation of
motion \eqref{fullveceq} for certain choices of $K^a {}_b {}^c {}_d$.
Consider a kinetic term for which $K^{(ab)c} {}_d = 0$.  This places
restrictions on the $\fc_i$ functions:
\begin{subequations}
  \label{maxfns}
  \begin{gather}
    \fc_1 + \fc_3 = 0  \\ \fc_4 = - \frac{1}{2} \fc_5 = \fc_6 \\ \fc_7
    = \fc_8 = 0
  \end{gather}
\end{subequations}
Alternately, this condition implies a kinetic term that can be
written in the form
\begin{multline}
  \label{Maxwellian}
  K^a {}_b {}^c {}_d \nabla_a A^b \nabla_c A^d \\ = \pm (\mathcal{H}_1
  g^{ac} + \mathcal{H}_2 A^a A^c)(\mathcal{H}_1 g^{bd} +\mathcal{H}_2
  A^b A^d) F_{ab} F_{cd}
\end{multline}
where $F_{ab} = 2 \nabla_{[a} A_{b]}$, $\fc_1 = \pm \mathcal{H}_1^2$,
and $\fc_4 = \pm \mathcal{H}_1 \mathcal{H}_2$.  (The signs here are
determined by the overall sign of $\fc_1$.)  As this kinetic term is
simply the familiar Maxwell field strength tensor contracted twice
with a ``generalized metric'' $\mathcal{H}_1 g^{ab} +\mathcal{H}_2 A^a
A^b$, we will call such kinetic terms (and theories containing them)
``pseudo-Maxwell.''

Taking the divergence of the vector equation of motion
\eqref{fullveceq} for a general $K^a {}_b {}^c {}_d$ and linearizing
about our chosen background, we find that
\begin{multline}
  \xi ( -f'_u \ba^a \nabla_a \delta R + f'_s \ba^b \ba^c
  \ba^a \nabla_a \delta R_{bc} + f_s \ba^b \nabla^a \delta
  R_{ba}) \\ - K^{ba} {}^c {}_d \delta( \nabla_a \nabla_b \nabla_c A^d)
  - 2 V''(\ba^2) \ba^a \ba_b \delta(\nabla_a A^b) = 0\end{multline}
For an arbitrary vector field $A^a$ and an arbitrary metric, we know
that 
\begin{multline}
  \label{thirdderid}
  \nabla_a \nabla_b \nabla_c A^d = \nabla_{(a} \nabla_{b)} (\nabla_c
  A^d) \\ {} + \frac{1}{2} \left( R_{abc} {}^e \nabla_e A^d - R_{abe} {}^d
    \nabla_c A^e \right).
\end{multline}
It can be then be seen that in the case $K^{(ba)c} {}_d = 0$, to
linear order in $\epsilon$ the divergence of the vector equation of
motion is simply
\begin{multline}
  \label{linvecdiv}
  \xi ( -f'_u \ba^a \nabla_a \delta R + f'_s \ba^b \ba^c
  \ba^a \nabla_a \delta R_{bc} + f_s \ba^b \nabla^a \delta
  R_{ba}) \\ = 2 V''(\ba^2) \ba^a \ba_b \delta(\nabla_a
  A^b) 
\end{multline}
(note that the quantity in brackets in equation \eqref{thirdderid} is
$\mathcal{O}(\epsilon^2)$.)  Using the linearized contracted Bianchi
identity $\nabla^a \delta R_{ab} = \frac{1}{2} \nabla_b \delta R$,
this last equation is equivalent to
\begin{multline}
  \ba^a \nabla_a \bigg( \xi \left( -f'_u + \frac{1}{2} f_s \right)
    \delta R + \xi f'_s \ba^b \ba^c \delta R_{bc} \\ {} -
  V''(\ba^2) \delta( A^2 ) \bigg)  = 0.
\end{multline}
where $\delta(A^2) = \delta(A^a A_a) = 2 \tilde{A}^a \ba_a + h_{ab}
\ba^a \ba^b$.  

This implies that in the case where $K^{(ab)c} {}_d$ = 0, if the
linearised quantity in brackets above vanishes on some hypersurface to
which $\ba^a$ is non-tangent, this quantity will vanish throughout
spacetime.  (Recall that $\ba^a$ is a constant vector field in
Minkowski space.) Thus, via an appropriate choice of boundary
conditions, we can impose
\begin{multline}
  \label{BCcond}
  \delta \mathcal{F} \equiv \xi \left( -f'_u + \frac{1}{2} f_s
  \right) \delta R + \xi f'_s \ba^b \ba^c \delta R_{bc} \\ - V''(\ba^2)
  \delta( A^2 ) = 0
\end{multline}
everywhere.\footnote{We have abused notation somewhat here, inasmuch
  as the quantity $\delta \mathcal{F}$ defined by \eqref{BCcond} is
  not obtained as the linearized variation of some quantity
  $\mathcal{F}$.  Nevertheless, we will continue to use $\delta
  \mathcal{F}$ throughout as a reminder that equations involving it
  are not exact, but only hold to linear order.}  This equation can be
interpreted as telling us how much the vector field moves ``up'' its
potential (recall that the value of the potential $V$ only depends on
$A^2$), and so we will call the equation \eqref{BCcond} the
``massive-mode'' condition.  When combined with the linearized vector
equations of motion \eqref{linveceq}, this yields
\begin{equation}
  \label{linveceqalt}
  \frac{1}{2} \delta (\mathcal{E}_A)_a = \xi f_s \ba^b \delta
  G_{ab} - K^b {}_a {}^c {}_d \delta (\nabla_b \nabla_c A^d) = 0.
\end{equation}

This massive-mode condition can then be used to impose further
conditions on $A^a$ and its derivatives.  It can be shown (see
Appendix \ref{ophypapp}) that by taking the appropriate combinations
of the derivatives of the equation of motion, we arrive at the
equation \label{Maxdyn}
\begin{equation}
  \label{linafseq}
  \mathfrak{O}_a {}^b [\ba^c \delta( \nabla_{[b} A_{c]})] = \xi f_s
  \ba^b \ba^c \partial_{[a} \delta G_{b]c}
\end{equation}
where $\mathfrak{O}_a {}^b$ is the \emph{flat-space} linear
second-order differential operator 
\begin{multline}
  \label{opdef}
  \mathfrak{O}_a {}^b \equiv \fc_1 \delta_a {}^b \Box \\ {}+ \fc_4
  \left( \delta_a {}^b \ba^c \ba^d \partial_c \partial_d +
    \ba^2 \partial_a \partial^b - \ba_a \ba^c \partial_c \partial^b
  \right).
\end{multline}

Thus, the operator $\mathfrak{O}_a {}^b$ applied to the one-form $v_a
\equiv \ba^b \delta( \nabla_{[a} A_{b]} )$ yields a quantity of order
$\xi$.  The properties of $\mathfrak{O}_a {}^b$ (see Appendix
\ref{ophypapp}) allow us to conclude that under the imposition of
appropriate boundary conditions, the quantity $v_a$ will itself be of
order $\xi$ as long as
\begin{equation}
  \fc_1 ( \fc_1 + \ba^2 \fc_4) > 0.
\end{equation}
Since we also have
\begin{equation}
\ba^a \delta (\nabla_b A_a) = \frac{1}{2} \delta( \nabla_b A^2) \sim
\mathcal{O}(\xi)
\end{equation}
from the massive-mode condition \eqref{BCcond} above, we can conclude
that under these assumptions, the quantity 
\begin{equation}
\ba^a \delta( \nabla_a A_b) = - 2 v_b + \frac{1}{2} \delta( \nabla_b
A^2) \sim \mathcal{O}(\xi)
\end{equation}
as well.  This condition, along with the massive-mode condition
\eqref{BCcond}, will become important in our analysis of the effective
gravitational equations below.

   
\section{Conditions on vector dynamics}
\label{applsec}

\subsection{The Einstein limit}
\subsubsection{General case}
\label{geneinsec}

Recall the second of Bailey and Kosteleck\'y's conditions above:
namely, that any Lorentz-violating corrections to the linearized
Einstein equation are linear in the parameter $\xi$.  This implies
that in the limit $\xi \to 0$, the equations of motion
\eqref{linEineq} and \eqref{linveceq} must together imply that the
conventional linearized Einstein equation is satisfied, i.e., that
$\delta G^{ab} = 0$.  In this limit, the equations of motion become
\begin{equation}
  \label{linEineqxi0}
  - \delta G^{ab} + \mathcal{Q}_K {}^{abc} {}_d {}^e \delta( \nabla_e
  \nabla_c A^d ) - V'' \ba^a \ba^b \delta (A^2) = 0,
\end{equation}
with $\mathcal{Q}_K$ defined as in \eqref{QKdef}, and
\begin{equation}
  \label{linveceqxi0}
  - K^b {}_a {}^c {}_d \delta (\nabla_b \nabla_c A^d) - V'' \ba_a
  \delta (A^2) = 0.
\end{equation}
We will further allow the functions $\fc_i(A^2)$ to be dependent on
$\xi$, defining functions $\mc_i(A^2)$ and $\md_i(A^2)$ such that
\begin{equation}
  \label{cddef}
  \fc_i = \mc_i + \xi \md_i + \mathcal{O}(\xi^2).
\end{equation}

For the two equations \eqref{linEineqxi0} and \eqref{linveceqxi0} to
imply the validity of the conventional linearized Einstein equation,
we must be able to eliminate the terms containing second derivatives
of the vector field from \eqref{linEineqxi0} using the vector equation
of motion \eqref{linveceqxi0}.  Since this must occur for an arbitrary
perturbation of the vector field, with arbitrary derivatives, we
conclude that this will only occur if for some tensor
$\mathcal{T}^{abf}$,
\begin{equation}
  \label{QKcond}
  \mathcal{Q}_K {}^{abc} {}_d {}^e = \mathcal{T}^{abf}  K^e
  {}_f {}^c {}_d
\end{equation}
in the limit $\xi \to 0$.  If this relation holds, then we can combine
the linearized Einstein equation and the linearized vector equation of
motion to obtain
\begin{equation}
  \label{lineinV}
  \delta G^{ab} = -V''(\ba^2) (\ba^a \ba^b +
  \mathcal{T}^{abc} \ba_c ) \delta(A^2).
\end{equation} 
This further implies that if the conventional Einstein equation is to
hold in the limit $\xi \to 0$, we must either have $\ba^a \ba^b +
\mathcal{T}^{abc} \ba_c = 0$ or $\delta (A^2) = 0$ in this limit.

What form must this tensor $\mathcal{T}^{abc}$ have?  For later
convenience, we will split it up into pieces of $\mathcal{O}(\xi^0)$
and $\mathcal{O}(\xi^1)$:
\begin{equation}
  \label{Tsplit}
  \mathcal{T}^{abc} = \mathcal{T}_0^{abc} + \xi \hat{\mathcal{T}}^{abc}.
\end{equation}
Moreover, since we are only concerned with the linearized equations,
we can take $\mathcal{T}^{abc}$ to be composed solely of background
quantities.  Since the only two geometric objects ``in play'' in the
background are the vector field $\ba^a$ and the flat metric
$\eta^{ab}$, and given the symmetry $\mathcal{T}^{abc} =
\mathcal{T}^{bac}$ inherent in the definition of $\mathcal{T}^{abc}$,
we conclude that $\mathcal{T}^{abc}$ must be of the form
\begin{equation}
  \label{T0formdef}
  \mathcal{T}_0^{abc} = U_1 \eta^{ab} \ba^c + U_2 \ba^{(a}
  \eta^{b)c} + U_3 \ba^a \ba^b \ba^c
\end{equation}
and
\begin{equation}
  \label{Thatformdef}
  \hat{\mathcal{T}}^{abc} = V_1 \eta^{ab} \ba^c + V_2 \ba^{(a}
  \eta^{b)c} + V_3 \ba^a \ba^b \ba^c,
\end{equation}
where the coefficients $U_i$ and $V_i$ can in principle be functions
of $\ba^2$.  Assuming that $\delta(A^2) \neq 0$, the
constraint that $\ba^a \ba^b + \mathcal{T}_0^{abc} \ba_c$ vanish
yields:
\begin{equation}
  \label{noVcond}
  U_1 = 0 \text{ and } U_2 + U_3 \ba^2 + 1 = 0.
\end{equation}

The question now becomes what form $K^a {}_b {}^c {}_d$ can have and
still satisfy the condition \eqref{QKcond}.  As with our other
quantities, we will split $K^a {}_b {}^c {}_d$ into
$\mathcal{O}(\xi^0)$ and $\mathcal{O}(\xi^1)$ parts:
\begin{equation}
  K^a {}_b {}^c {}_d = (K_0)^a {}_b {}^c {}_d + \xi \hat{K}^a {}_b
  {}^c {}_d + \mathcal{O}(\xi^2).
\end{equation}
Note that due to the decomposition \eqref{cddef}, $(K_0)^a {}_b {}^c
{}_d$ or $\hat{K}^a {}_b {}^c {}_d$ can be obtained by taking the
original definition \eqref{Kdef} of $K^a {}_b {}^c {}_d$ and replacing
$\fc_i$ by $\mc_i$ or $\md_i$, respectively.  Similarly, we will
define 
\begin{equation}
  \label{Qsplit}
  \mathcal{Q}_K {}^{abc} {}_d {}^e = (\mathcal{Q}_{K0}) {}^{abc} {}_d {}^e
  + \xi \hat{\mathcal{Q}}_K {}^{abc} {}_d {}^e + \mathcal{O}(\xi^2).
\end{equation}

In the limit $\xi = 0$, we thus have the condition
\begin{equation}
  (\mathcal{Q}_{K0}) {}^{abc} {}_d {}^e = \mathcal{T}_0^{abf} (K_0)^e
  {}_f {}^c {}_d 
\end{equation}
Both sides of this equation consist of various five-index tensors
constructed from $\ba^a$ and the metric, with various coefficients
given in terms of $U_2$ and the $\mc_i$ functions.  (Their exact forms
are given in Appendix \ref{nastytensors}, Equations \eqref{Q0full} and
\eqref{TK0full}.)  Matching these coefficients, we obtain a set of
eleven equations which the $\mc_i$ functions and $U_2$ must satisfy.
(We of course want a non-trivial solution for the $\mc_i$
coefficients.)  Examination of the resulting equations shows that we
must have $U_2 = -2$ and $U_3 = \ba^{-2}$, and that the functions
$\mc_i$ must satisfy
\begin{equation}
\mc_1 = - \mc_3 = -\bar{A}^2 \mc_4 = \frac{1}{2} \ba^2 \mc_5 \text{
  and } \mc_7 = 0
\end{equation}
with $\mc_6$ and $\mc_8$ arbitrary.  This implies a vector kinetic
term that can be rewritten in the form
\begin{multline}
  \label{weirdass}
  K^a {}_b {}^c {}_d \nabla_a A^b \nabla_c A^d \\ = \mathcal{G}_1 (g^{ac}
  - A^{-2} A^a A^c) (g^{bd} 
  - A^{-2} A^b A^d ) F_{ab} F_{cd} \\ + \left( \mathcal{G}_2
    g^{ab} + 
    \mathcal{G}_3 A^a A^b \right) \nabla_a (A^2) \nabla_b (A^2)
\end{multline}
where $F_{ab} = 2 \nabla_{[a} A_{b]}$ and the coefficients
$\mathcal{G}_i$ are functions of $A^2$, related to the $\mc_i$
functions by $\mc_1 = 2 \mathcal{G}_1$, $\mc_6 = 4 \mathcal{G}_2 - 2
A^{-2} \mathcal{G}_1$, and $\mc_8 = 4 \mathcal{G}_3$.

\subsubsection{Pseudo-Maxwell dynamics}

In the previous subsection, we assumed that a general form for $K^a
{}_b {}^c {}_d$.  However, as was noted at the end of Section
\ref{specialKsec}, a ``pseudo-Maxwell'' vector kinetic term,
satisfying $K^{(ab)c} {}_d = 0$, will behave somewhat differently.
The linearized solutions obtained from such an action will, with the
imposition of appropriate boundary conditions, also meet additional
self-consistency conditions due to properties of the linearized
equations of motion. In particular, in the $\xi \to 0$ limit, the
condition \eqref{BCcond} becomes
\begin{equation}
  V''(\ba^2) \delta(A^2) = 0.
\end{equation}
This allows us to ignore the constraints \eqref{noVcond} on
$\mathcal{T}^{abc}$, as they were imposed by the requirement that the
right-hand side of Equation \eqref{lineinV} vanish.  We therefore only
have the requirement that the second derivatives of $A^a$ vanish, as
expressed by \eqref{QKcond}, in order to obtain a valid Einstein
limit.  In this case, the full tensors are given by Equations
\eqref{Q0maxfull} and \eqref{TK0maxfull} in Appendix
\ref{nastytensors}.  Once again, we perform the
matching of coefficients between these two tensors, yielding a set of
equations that must be satisfied by the $\mc_i$ and $U_i$ functions.
Assuming that $\mc_1 \neq -\ba^2 \mc_4$, these two tensors will be
equal if and only if $U_2 = -2$ and $U_1 = U_3 = 0$.\footnote{Note
  that the case where $\mc_1 = - \ba^2 \mc_4$ is a special case of the
  kinetic term \eqref{weirdass} derived in the previous section.}  We
have thus found two possible vector field kinetic terms, given by
\eqref{Maxwellian} and \eqref{weirdass}, for which the conventional
Einstein limit is recovered in the limit of no direct coupling to
curvature.

\subsection{Adding Lorentz violation}

In the above section, we obtained vector actions which satisfied
Condition 2 above; namely, in the limit of no direct coupling to
curvature, these actions yielded linearized equations of motion that
implied the conventional linearized Einstein equation $\delta G^{ab} =
0$.  We now wish to ``turn on'' direct coupling between the curvature
and the vector field by setting $\xi \neq 0$ and place further
constraints on the form of these actions.

Although Condition 2 does not yield any constraints on the form of the
equations of motion at $\mathcal{O}(\xi)$, we can still constrain the
vector action by imposing Condition 5: we must be able to eliminate
the derivatives of $A^a$ from the metric equation of motion
\eqref{linEineq} via use of the vector equation of motion
\eqref{linveceq}.  In particular, the terms in \eqref{linEineq} which
contain derivatives of the vector field can be written in the form
\begin{multline}
  (\mathcal{Q}_K {}^{abc} {}_d {}^e + \xi \mathcal{Q}_R {}^{abc} {}_d
  {}^e) 
  \delta( \nabla_e \nabla_c A^d) \\ = \left( (\mathcal{Q}_{K0}) {}^{abc}
    {}_d {}^e + \xi (\hat{\mathcal{Q}}_K {}^{abc} {}_d {}^e +
    \mathcal{Q}_R {}^{abc} {}_d {}^e) \right) \delta( \nabla_e
  \nabla_c A^d)  
\end{multline}
Using the vector equation of motion \eqref{linEineq} and the
condition \eqref{QKcond}, we can rewrite this as
\begin{multline}
  \label{xiderivs}
  (\mathcal{Q}_K {}^{abc} {}_d {}^e + \xi \mathcal{Q}_R {}^{abc} {}_d
  {}^e) 
  \delta( \nabla_e \nabla_c A^d) \\ \simeq \xi \left( - \mathcal{T}_0^{abf}
    \hat{K}^e {}_f {}^c {}_d + \hat{\mathcal{Q}}_K {}^{abc} {}_d {}^e +
    \mathcal{Q}_R {}^{abc} {}_d {}^e \right) \delta( \nabla_e
  \nabla_c A^d) 
\end{multline}
where the ``$\simeq$'' symbol here means ``up to terms not involving
derivatives of $A^a$.''  We can further simplify this expression by
noting that in an arbitrary spacetime, 
\begin{equation}
  \nabla_a \nabla_b A^c = \nabla_{(a} \nabla_{b)} A^c - \frac{1}{2}
  R_{abd} {}^c A^d 
\end{equation}
or, in our case, 
\begin{equation}
  \delta( \nabla_a \nabla_b A^c ) = \delta( \nabla_{(a} \nabla_{b)}
  A^c ) - \frac{1}{2} \delta R_{abd} {}^c \ba^d 
\end{equation}
up to linear order in $\epsilon$.  Thus, at $\mathcal{O}(\xi)$ we only
need to eliminate the symmetrized second derivatives from the metric
equation of motion \eqref{linEineq}; the antisymmetrized second
derivatives will merely result in contractions of $\ba^a$ with the
linearized Riemann tensor, which are expected if the effective
linearized gravitational equation is to be of the form
\eqref{effgrav}.  This will occur if $\hat{\mathcal{T}}^{abc}$ (the
$\mathcal{O}(\xi)$ contribution to $\mathcal{T}^{abc}$ defined in
\eqref{Tsplit}) satisfies the equation
\begin{equation}
  \label{Qcondxi}
  \mathcal{Q}_R {}^{ab(c} {}_d {}^{e)} + \hat{\mathcal{Q}}_K
  {}^{ab(c} {}_d {}^{e)} = \hat{\mathcal{T}}^{abf} K^{(e} {}_f {}^{c)}
  {}_d + \mathcal{T}_0^{abf} \hat{K}^{(e} {}_f {}^{c)}
  {}_d.
\end{equation}
This equation is essentially the $\mathcal{O}(\xi)$ analog of Equation
\eqref{QKcond}.

We can now proceed with the analysis of this equation as we did in the
$\xi = 0$ limit: we write out the left-hand and right-hand sides in
terms of various five-index tensors constructed from $\eta^{ab}$ and
$\ba^a$, and match coefficients to determine the possible forms of the
$\md_i$'s and their corresponding $\mathcal{T}^{abc}$ tensors.
Expressions for the resulting tensors are given in Appendix
\ref{nastytensors}; the left-hand side of \eqref{Qcondxi} is given by
equation \eqref{tensmatch1}, while the right-hand side is given by
\eqref{tensmatch2}.

\subsubsection{General case}

In the case where $K^{(ab)c} {}_d \neq 0$, we found in Section
\ref{geneinsec} that the kinetic terms for the vector must be given by
\eqref{weirdass}, with $U_3 = \ba^{-2}$.  We now wish to match the
coefficients in \eqref{tensmatch1} and \eqref{tensmatch2} to see what
conditions can be placed on the $\md_i$ coefficients and the functions
$f_s$ and $f_u$.  Substituting in the appropriate relations for the
$\mc_i$'s and $U_3$, we find that if \eqref{tensmatch1} and
\eqref{tensmatch2} are to agree, we are forced to set
\begin{equation}
  f'_u(A^2) = 0
\end{equation}
and
\begin{equation}
  f_s(A^2) = 0.
\end{equation}
These conditions can most easily be seen from the coefficients of
$\eta^{e(a} \eta^{b)c} \ba_d$ and $\eta^{ab} \ba^{(c} \delta^{e)}
{}_d$, respectively.  In other words, the vector model whose kinetic
term is given by \eqref{weirdass} cannot be modified with a
Lorentz-violating curvature coupling of the form \eqref{genericLV} and
still satisfy the assumptions of the Bailey-Kosteleck\'y formalism.
(Note that setting $f_u(A^2)$ to a non-zero constant merely changes
the effective value of $G$.)  Thus, this theory cannot be successfully
be analyzed under this formalism unless Lorentz-violating effects
induced by the coupling term $\mathcal{L}_{LV}$ vanish.

\subsubsection{Pseudo-Maxwell dynamics}
\label{MaxLV}

The obvious next step is to attempt the same coefficient matching for
pseudo-Maxwell vector theories, as defined in \eqref{Maxwellian}.
However, when we na\"ively do so, we find that the same logic that
forced us to abandon Lorentz violation in the vector model
\eqref{weirdass} again forces the Lorentz-violating functions $f_u$
and $f_s$ to vanish in the case of pseudo-Maxwell kinetic terms.  This
stands in opposition to the fact Bailey and Kosteleck\'y successfully
applied their formalism to the so-called ``bumblebee model''
\cite{Kosgrav} in their original paper \cite{BK}; the kinetic term for
this model is the same as our pseudo-Maxwell kinetic term in the
special case $\fc_1 = \text{constant}$ and $\fc_4 = 0$.  What have we failed to take into
account?

The missing pieces are the conditions on the linearized derivatives of
$A^a$ derived in Section \ref{Maxdyn}.  Namely, we found that under
the imposition of certain boundary conditions, we have
\begin{equation}
  \label{Maxderconds}
  \ba^a \delta(\nabla_a A_b) \sim \ba^a \delta(\nabla_b A_a) \sim
  \mathcal{O}(\xi)
\end{equation} 
everywhere in the spacetime.  The role of these conditions is easiest
to see by returning to Equation \eqref{xiderivs} and examining the
$\mathcal{O}(\xi^1)$ derivative terms remaining in the equations of
motion after eliminating the $\mathcal{O}(\xi^0)$ derivative terms.
To wit, suppose there exist tensors $\hat{\mathcal{T}}^{abf}$,
$c^{abc} {}_d$, $\tilde{c}^{abc} {}_d$, and $d^{abcd}$ such that we
can write
\begin{multline}
  \label{ccddef}
  - \mathcal{T}_0^{abf}
  \hat{K}^e {}_f {}^c {}_d + \hat{\mathcal{Q}}_K {}^{abc} {}_d {}^e +
  \mathcal{Q}_R {}^{abc} {}_d {}^e \\ = \hat{\mathcal{T}}^{abf} (K_0)^e {}_f
  {}^c {}_d + c^{abc} {}_d \ba^e + \tilde{c}^{abe}{}_d \ba^c +
  d^{abce} \ba_d 
\end{multline}
The conditions \eqref{Maxderconds} on the derivatives of $A^a$ imply
that to linear order in $\epsilon$, $\ba^c \delta(\nabla_e \nabla_c
A^d)$ and $\ba_d \delta(\nabla_e \nabla_c A^d)$ are of order $\xi$;
similarly, to this order in $\epsilon$ we will have
\begin{align}
  \ba^e \delta(\nabla_e \nabla_c A^d) &= \ba^e \delta(\nabla_c \nabla_e
  A^d) + \ba^e \delta R_{cef} {}^d \ba^f \notag \\ & = \ba^e \ba^f \delta R_{cef}
  {}^d + \mathcal{O}(\xi).
\end{align}
Thus, if Equation \eqref{ccddef} holds, we will have
\begin{multline}
  \label{Maxderelim}
  \xi( - \mathcal{T}_0^{abf} \hat{K}^e {}_f {}^c {}_d +
  \hat{\mathcal{Q}}_K {}^{abc} {}_d {}^e + \mathcal{Q}_R {}^{abc} {}_d
  {}^e ) \delta(\nabla_e \nabla_c A^d) \\ = \xi(\hat{\mathcal{T}}^{abf}
  (K_0)^e {}_f {}^c {}_d \delta(\nabla_e \nabla_c A^d) + c^{abc} {}_d
  \ba^e \ba^f R_{cef} {}^d) \\ + \mathcal{O}(\xi^2), 
\end{multline}
since all the other terms on the right-hand side of \eqref{ccddef} are
of $\mathcal{O}(\xi)$ when contracted with $\delta(\nabla_e \nabla_c
A^d)$.\footnote{Note that the decomposition in \eqref{ccddef} is
  ambiguous: it does not address what is to be done with terms of the
  form $C^{abc} \ba_d \ba^e$, for instance.  However, it is easily
  seen from \eqref{Maxderelim} that such terms will vanish when
  contracted with the Riemann tensor, so it does not matter whether we
  consider them to be part of $c^{abc}{}_d$ or $d^{abce}$.}  In
essence, the derivative conditions \eqref{Maxderconds} allow us to
``ignore'' certain of the equations arising from the
coefficient-matching implicit in \eqref{Qcondxi} at a given order in
$\xi$.

To perform this decomposition, we first note that by taking the
equation $\mathcal{T}_0^{abf} (K_0)^e {}_f {}^c {}_d =
(\mathcal{Q}_{K0})^{abc} {}_d {}^e$ and replacing the $\mc_i$
functions with $\md_i$ functions, we obtain
\begin{equation}
  \mathcal{T}_0^{abf} \hat{K}^e {}_f {}^c {}_d =
  \hat{\mathcal{Q}}_K {}^{abc} {}_d {}^e.
\end{equation}
(To put this another way, the relations \eqref{maxfns} hold to all
orders in $\xi$, and so $\mathcal{T}_0^{abf} K^e {}_f {}^c {}_d =
\mathcal{Q}_K {}^{abc} {}_d {}^e$ to all orders.)  Thus, the first two
terms on the left-hand side of \eqref{ccddef} cancel, and we merely
need to examine $\mathcal{Q}_R {}^{abc} {}_d {}^e$ to find out the
required form of the tensors on the right-hand side.  The form of
$\mathcal{Q}_R {}^{abc} {}_d {}^e$ is given by \eqref{QRdef}; for a
$\hat{\mathcal{T}}^{abf}$ given by \eqref{Thatformdef}, the quantity
$\hat{\mathcal{T}}^{abf} K^e {}_f {}^c {}_d$ is given by
\begin{multline}
  \hat{T}^{abf} K^e {}_f {}^c {}_d = V_2 \mc_1 ( \ba^{(a} \delta^{b)}
  {}_d \eta^{ce} - \ba^{(a} \eta^{b)c} \delta^e {}_d ) \\ + V_1 (\mc_1 +
  \ba^2 \mc_4)(\eta^{ab} \eta^{ce} \ba_d - \eta^{ab} \delta^e {}_d
  \ba^c ) \\
  + (V_2 \mc_4 + V_3 (\mc_1 + \ba^2 \mc_4) ) \ba^a \ba^b
  (\eta^{ce} \ba_d - \ba^c \delta^e {}_d) \\ + V_2 \mc_4 ( \ba^{(a}
  \delta^{b)} {}_d \ba^c \ba^e - \ba^{(a} \eta^{b)c} \ba_d \ba^e ).
\end{multline}
Comparing these equations, we can then see that Equation
\eqref{ccddef} is satisfied if $\hat{\mathcal{T}}^{abc}$ has
\begin{equation}
  V_2 \mc_1 = - f_s,
\end{equation}
with $V_1$ and $V_3$ arbitrary, and
\begin{equation}
  c^{abc} {}_d = f_s \left( - \frac{1}{2} \eta^{ab} \delta^c {}_d +
    \eta^{c(a} \delta^{b)} {}_d \right).
\end{equation}
Note that this latter quantity is independent of the form of
$\hat{\mathcal{T}}^{abf}$. 

Finally, we confirm that the effective gravitational equations are of
the proper form for these pseudo-Maxwell models.  Applying the
massive-mode condition \eqref{BCcond} to the linearized Einstein
equation \eqref{linEineq}, we obtain 
\begin{multline}
  \delta G^{ab} = \xi \left( f_u \delta G^{ab} - \frac{1}{2} f_s
    \left( \ba^a
    \ba^b \delta R - \eta^{ab} \ba^c \ba^d \delta
    R_{cd} \right) \right) \\ + ( \mathcal{Q}_K {}^{abc} {}_d {}^e + \xi
  \mathcal{Q}_R {}^{abc} {}_d {}^e ) \delta( \nabla_e \nabla_c A^d ).
\end{multline}
Using the linearized vector equation of motion \eqref{linveceqalt}
contracted with $\mathcal{T}_0^{abf} = - 2 \ba^{(a} \eta^{b)f}$, we
can eliminate the $\mathcal{O}(\xi^0)$ derivative terms to obtain
\begin{multline}
  \delta G^{ab} = \xi \bigg( f_u \delta G^{ab} + \frac{1}{2} f_s
  \ba^a \ba^b \delta R - 2 f_s \ba^{(a} \delta R^{b)} {}_c \ba^c \\ {}
  +
  \frac{1}{2} f_s \eta^{ab} \ba^c \ba^d \delta R_{cd} +
  \mathcal{Q}_R {}^{abc} {}_d {}^e \delta(\nabla_e \nabla_c A^d )
  \bigg).
\end{multline}
Lastly, the remaining derivatives of $A^a$ in the above equation can
be eliminated using the derivative conditions, as noted above in
equation \eqref{Maxderelim};  this yields 
\begin{multline}
  \delta G^{ab} = \xi \bigg( f_u \delta G^{ab} + f_s \bigg(
      \frac{1}{2} \ba^a \ba^b \delta R - 2 \ba^{(a} \delta R^{b)}
      {}_c \ba^c \\ + \eta^{ab} \ba^c \ba^d \delta R_{cd} + \ba^c \ba^d
      \delta R^a {}_{cd} {}^b 
    \bigg) \bigg).
\end{multline}
In our parametrization, the bumblebee model \cite{Kosgrav} is
obtained by setting $f_s = 1$ and $f_u = 0$.  Plugging in these
values, this effective equation for $\delta G^{ab}$ reduces to the
form of the effective gravitational equation \eqref{sueffgrav} found
by Bailey and Kosteleck\'y, with an ``effective $\bar{u}$'' of
$-\frac{3}{4} \ba^2$ and with $\bar{s}^{ab} = \ba^a \ba^b -
\frac{1}{4} \eta^{ab} \ba^2$.

\section{Discussion}
\label{discsec}

We have systematically examined the dynamics of vector-tensor gravity
theories with spontaneous Lorentz symmetry breaking.  The primary
constraints on the form of these theories were obtained by imposing
two of Bailey \& Kosteleck\'y's conditions: First, we required that
the equations have the correct weak-field Einstein limit $\delta
G_{ab} = 0$ when the Lorentz-violating terms \eqref{genericLV} are
``turned off'' (Condition 2 of the list in Section \ref{condlist});
second, we required that the linearized stress-energy of the vector
field vanish automatically when the linearized vector equations of
motion held (Condition 5).  The first of these requirements led us to
the conclusion that the kinetic terms for our vector fields must be of
the form \eqref{Maxwellian} or \eqref{weirdass}.  The vanishing of the
linearized vector stress-energy was found to be a somewhat more subtle
issue; we found that under the imposition of appropriate boundary
conditions, the so-called pseudo-Maxwell vector models (those with
kinetic terms of the form \eqref{Maxwellian}) could lead to effective
gravitational equations expressed solely in terms of the metric.

It is important to reiterate that the imposition of boundary
conditions is necessary to obtain effective gravitational equations of
the form used by Bailey and Kosteleck\'y in their post-Newtonian
analysis; as was noted at the beginning of Section \ref{MaxLV}, an
arbitrary solution of the vector equations of motion will not have the
proper relations between the derivatives of the vector field to cause
the linearized vector stress-energy to vanish.  In a certain sense,
this confirms the aptness of the name ``bumblebee model''.  This name
was originally inspired by the notion that according to received
wisdom, bumblebees should not be able to fly; na\"ive calculations by
engineers and entomologists in the 1930s seemed to show that the
bumblebee's wings were too small to allow it to fly, and only once
more subtle aerodynamic effects were taken into account was the
mystery explained.  Similarly, a na\"ive comparison of the bumblebee
vector equations of motion with its stress-energy causes us to
conclude that we cannot introduce Lorentz-violating gravitational
effects into the model; only once more subtle effects (namely, proper
boundary conditions) are taken into account can Lorentz violation in
the bumblebee model ``fly.''

This said, the technique of imposing boundary conditions to obtain the
desired effective gravitational equations is not entirely rigourous.
In particular, we used the somewhat vague statement that ``solutions
depend continuously on initial data'' to argue that the quantity
$\ba^a \delta (\nabla_a A_b)$ was of order $\xi$.  While this is true,
the notion of continuity associated with well-posedness of an initial
value problem is defined in terms of the norms of the solutions on
certain Sobolev spaces, and is not easy to gain a simple intuition
about (see Chapter 10 of \cite{Wald}).  The notion of ``continuous
dependence on initial data'' (and, by Duhamel's principle, on sources)
does allow us to say that we can always make $\ba^a \delta (\nabla_a
A_b)$ as small as we like by tuning $\xi$ to be ``sufficiently
small''; however, it is far from clear how small is ``sufficient.''
It would be instructive to obtain more careful estimates of how
critically the magnitude of $\ba^a \delta (\nabla_a A_b)$ depends on
$\xi$; however, such an analysis is well outside the scope of this
paper.

In some sense, the fact that only pseudo-Maxwell kinetic terms are
acceptable for Lorentz violation is not entirely surprising given the
Bailey-Kosteleck\'y formalism's requirement of cancellations in the
equations of motion.  The quantity $\nabla_a A_b$ will, in general,
depend both on derivatives of the vector field and derivatives of the
metric (this latter dependence can be thought of as arising from the
Christoffel symbols implicit in $\nabla_a A_b$.)  A vector kinetic
term containing an arbitrary contraction of $\nabla_a A_b$ with itself
and other fields will then, in general, lead to a ``cross term''
between derivatives of the vector and derivatives of the metric in the
kinetic terms of the theory \cite{IsenNest}. However, the
antisymmetrized derivative $\nabla_{[a} A_{b]}$ is independent of the
metric, and so the kinetic terms for the metric and the vector will be
decoupled when we contract $\nabla_{[a} A_{b]}$ with itself.  It is
therefore not surprising that this special property should have some
bearing on the relation between the vector equations of motion and the
gravitational equations of motion.

In the case of $\fc_4 = 0$ and $\fc_1$ constant, the pseudo-Maxwell
theories we have been discussing become a simple Maxwell action for
the vector field (albeit without gauge symmetry, which is broken by
the presence of the potential.)  However, the theories for which
$\fc_4 \neq 0$ do not appear to have been previously considered in the
literature, at least as far as concerns Lorentz-violating effects.  In
some sense, the presence of a $\fc_4 \neq 0$ term causes Lorentz
violation for the Lorentz-violating field itself: at the linearized
level, small perturbations of the vector field ``see'' the effective
metric $\mathcal{H}_1 g^{ab} + \mathcal{H}_2 A^a A^b$ (as defined in
\eqref{Maxwellian}), rather than the spacetime metric $g^{ab}$.  In
particular, in the bumblebee model the Nambu-Goldstone modes of the
Lorentz-violating vector field can be interpreted as a Maxwell field
in a particular gauge \cite{LVphoton}.  If we na\"ively extended this
interpretation to a general pseudo-Maxwell theory, one would expect
that the ``speed of light'' would be different from the ``speed of
gravity'', as the two fields would propagate on the null cones of two
different metrics.  Under such an interpretation the ``photon'' would
almost certainly propagate anisotropically; it is also possible that
such an interpretation would predict vacuum birefringence.
Experimental bounds on such phenomena could then place bounds on the
relative values of $\mathcal{H}_1$ and $\mathcal{H}_2$.  That said,
this intuitional understanding may be complicated by the fact that the
correspondence in the above-mentioned work \cite{LVphoton} is in a
non-standard gauge.  It is also known that this correspondence does
not carry over to theories with more general kinetic terms than the
bumblebee model \cite{genLVham}, though the class of models examined
in this last work did not include the pseudo-Maxwell theories we have
found.  More work is needed to elucidate the correspondence (if any)
between Maxwell theory and the Nambu-Goldstone modes of these new
theories.

Finally, it is important to note that our results imply that the
Bailey-Kosteleck\'y formalism cannot successfully analyze theories
with non-standard kinetic terms \cite{einaeth,TeVeS,instaeth,sigaeth}.
This does not imply that post-Newtonian effects in such theories
cannot be analyzed; in fact, Bailey and Kosteleck\'y did precisely
this in their original paper \cite{BK} for a Lagrangian identical to
what Carroll \textit{et al.}\ later called sigma-{\ae}ther theory
\cite{sigaeth}.  It is further possible that such a theory might in
fact provide a viable model of Lorentz violation, consistent with
current experimental constraints, even though it does not fit into the
Bailey-Kosteleck\'y formalism.  In the absence of a more general
formalism for gravitational Lorentz violation, however, such theories
will have to be analyzed on a case-by-case basis.

\begin{acknowledgments}
  I would like to thank Alan Kosteleck\'y for useful discussions
  leading to this work.  This work was supported in part by the United
  States Department of Energy, under Grant No.~DE-FG02-91ER40661.
\end{acknowledgments}

\appendix
\begin{widetext}
\section{Derivation and hyperbolicity of the operator $\mathfrak{O}_a
  {}^b$} 
\label{ophypapp}

Consider the following linearized combination of the vector equations
of motion:
\begin{multline}
  \label{eqfst}
  \frac{1}{2} \delta \left( A^b \left(\nabla_b (\mathcal{E}_A)_a -
      \nabla_a (\mathcal{E}_A)_b \right) \right)
  = - \ba^f K^{dec} {}_{[a} \delta^b {}_{f]} \delta( \nabla_b \nabla_c
  \nabla_d A_e ) - V''(\ba^2) \ba^b \ba_{[a} \nabla_{b]} \delta(A^2)
  \\ + \xi \left( - f'_u \ba^b \ba_{[a} \nabla_{b]} \delta R + f'_s
    \ba^b \ba^c \ba^d \ba_{[a} \nabla_{b]} \delta R_{cd} - f_s \ba^b
    \ba^c \nabla_{[a} \delta R_{b]c} \right) 
\end{multline}
Writing out the term $- \ba^f K^{dec} {}_{[a} \delta^b {}_{f]}$ for
a theory in which $K^{(ab)c} {}_d = 0$, we find
\begin{equation}
- \ba^f K^{dec} {}_{[a} \delta^b {}_{f]} = \fc_1 ( 3 \ba^{[b}
\delta^d {}_a \eta^{e]c} + \delta^{[d} {}_a \ba^{e]} \eta^{bc} ) +
\fc_4 (\ba^b \ba^c \delta^{[d} {}_a \ba^{e]} + \ba^2 \delta^b {}_a
\eta^{c[d} \ba^{e]} - \ba_a \ba^b \eta^{c[d} \ba^{e]}  )
\end{equation}
Since $\delta( \nabla_{[b} \nabla_{c]} \nabla_d A_e ) \sim
\mathcal{O}(\epsilon^2)$ and $\nabla_{[a} \nabla_b A_{c]} = 0$, we can
rewrite the first term on the right-hand side of \eqref{eqfst} (to
linear order) as
\begin{equation}
  - 2 \ba^f K^{dec} {}_{[a} \delta^b {}_{f]} \delta( \nabla_b \nabla_c
  \nabla_d A_e ) = -2 \ba^f K^{dec} {}_{[a} \delta^b {}_{f]} \delta(
  \nabla_c \nabla_b \nabla_d A_e ) = 2 \mathfrak{O}_a {}^d [\ba^e
  \delta( \nabla_{[d} A_{e]})] 
\end{equation}
\end{widetext}
Further, applying the massive-mode condition $\delta \mathcal{F}=0$,
we can eliminate the term proportional to $V''(\ba^2)$ from
\eqref{eqfst}, yielding
\begin{equation}
  \label{linafseqapp}
  \mathfrak{O}_a {}^b [\ba^c \delta( \nabla_{[b} A_{c]})] = \xi f_s
  \ba^b \ba^c \partial_{[a} \delta G_{b]c}
\end{equation}
when the linearized vector equation of motion is satisfied.

Thus, the quantity $v_a \equiv \ba^b \delta( \nabla_{[a} A_{b]})$ will
satisfy a second-order differential equation \eqref{linafseqapp} in
flat spacetime.  Moreover, the source for this equation is ``small'',
i.e., of order $\xi$.  We are thus led to the following question:
under what conditions will the solution for $v_a$ itself be of order
$\xi$?  More precisely, let us pick some time coordinate $t$ on
Minkowski space.  We know that if we set $\xi = 0$, $v_a = 0$ for all
$t$ is a valid solution of the Cauchy problem for \eqref{linafseqapp}
with the boundary condition $v_a(t_0) = 0$ and $\partial v_a/\partial
t |_{t_0} = 0$.  We wish to know whether, as we ``tune'' $\xi$ to
zero, the solutions of $v_a$ go ``smoothly'' to zero for these
boundary conditions.

This is precisely the question of whether the operator $\mathfrak{O}_a
{}^b$ has a well-posed initial-value formulation.\footnote{Note that a
  ``small'' variation in the source terms in \eqref{linafseqapp} can
  be mapped to a ``small'' variation in the boundary conditions via
  Duhamel's principle.}  While the general problem of whether an
arbitrary operator possesses an initial-value formulation can be quite
subtle, for operators in flat spacetime with constant coefficients
(such as $\mathfrak{O}_a {}^b$) the situation is more clear-cut.
Suppose $\mathfrak{O}_a {}^b$ is a linear $m^\text{th}$-order
differential operator which operates on $N$-tuples of functions in
flat spacetime.  (Thus, an equation of the form $\mathfrak{O}_a {}^b
v_b = 0$ is a system of $N$ linear $m^\text{th}$-order differential
equations.)  Associated with any such operator we can find an $N
\times N$ polynomial-valued matrix $P_a {}^b (\lambda,\vec{\zeta})$
such that
\begin{equation}
  \label{Pdef}
  P_a {}^b \left( \frac{\partial}{\partial t}, \vec{\nabla} \right) =
  \mathfrak{O}_a {}^b,
\end{equation}
i.e., if we take $P_a {}^b$ and replace $\lambda$ by
$\partial/\partial t$ and $\vec{\zeta}$ by $\vec{\nabla}$, we obtain
the operator $\mathfrak{O}_a {}^b$.  We will further assume that the
matrix $P_a {}^b$ is constant with respect to space and time.  It can
then be shown \cite{Johnbook,Johnart} that such an operator has a
well-posed initial value formulation (with respect to an initial-data
surface $t = \text{constant}$) if and only if there exists a
real number $c$ such that the $mN$ roots $\lambda_i$ of the equation
\begin{equation}
  \label{gardingpoly}
  \det \left[ P( i \lambda, i \vec{\zeta} ) \right] = 0
\end{equation}
satisfy $\Im(\lambda_i) > -c$ for all real vectors $\vec{\zeta}$.
Such an operator is said to be ``hyperbolic in the sense of
G\r{a}rding.''

To apply this result to the case of the operator $\mathfrak{O}_a
{}^b$, let us choose a Cartesian coordinate system on flat spacetime
$\{t,x,y,z \}$ for which $\ba^x = \ba^y = 0$.  Then the polynomial
defined by \eqref{gardingpoly} becomes
\begin{multline}
(\fc_1 + \ba^2 \fc_4)(\lambda^2 - \vec{\zeta}^2) \\ {} \times \left( \fc_1
  (\lambda^2 - \vec{\zeta}^2) - \fc_4 (\ba^t \lambda + \ba^z \zeta_3
  )^2 
\right)^3 = 0
\end{multline}
This polynomial has roots when $\lambda_i^2 = \vec{\zeta}^2$ due to
its second factor; these will obviously have $\Im(\lambda_i) = 0$ for
all real $\vec{\zeta}$.  The third factor, meanwhile, is a slightly
more complicated quadratic polynomial in $\lambda$; its roots can be
shown to be real if its discriminant is positive:
\begin{equation}
  \mathfrak{D} \equiv \fc_1 \left( \left( \fc_1 - \fc_4 \left(\ba^t
      \right)^2 
    \right) \zeta_\perp^2 + \left( \fc_1 + \fc_4 \ba^2
    \right) 
    \zeta_z^2 \right) 
  > 0,
\end{equation}
where $\zeta_\perp^2 \equiv \zeta_x^2 + \zeta_y^2$.  If the quantity
$\mathfrak{D}$ is negative for some value of $\vec{\zeta}$, the
imaginary part of these roots will be $\pm \sqrt{\mathfrak{D}}$.
Moreover, should this quantity $\mathfrak{D}$ be negative for some
real vector $\vec{\zeta}$, the magnitude of the imaginary part of
these roots can be made arbitrarily large: if $\Im(\lambda_i) = \pm
\sqrt{\mathfrak{D}_0}$ for a given $\vec{\zeta} = \vec{\zeta}_0$, then
$\Im(\lambda_i) = \pm M \sqrt{\mathfrak{D}_0}$ for $\vec{\zeta} = M
\vec{\zeta_0}$.  Thus, the operator $\mathfrak{O}_a {}^b$ defined in
\eqref{opdef} will be hyperbolic in the sense of G\r{a}rding if and
only if $\mathfrak{D}$ is a positive definite quadratic form in
$\vec{\zeta}$, i.e., if
\begin{equation}
  \fc_1 (\fc_1 - \fc_4 \left(\ba^t \right)^2 ) > 0 \text{ and } \fc_1
  (\fc_1 + \fc_4 \ba^2 ) > 0.
\end{equation}

We can therefore conclude that in any frame in which these
inequalities hold, we can then impose boundary conditions on some
initial-time surface $t = t_0$ such that $\ba^b \delta( \nabla_{[a}
A_{b]}) \sim \mathcal{O}(\xi)$ throughout the spacetime.  We can
further ask that such a frame have $A^t \neq 0$; if this is the case,
then the massive-mode condition \eqref{BCcond} can also be imposed on
the surface $t = t_0$, and it will follow (via the linearized
equations of motion) that the massive-mode condition is satisfied
everywhere.  Such a frame will necessarily exist if 
\begin{equation}
  \fc_1 (\fc_1 + \ba^2 \fc_4 ) > 0.
\end{equation}
(If $\ba^2 < 0$, the frame in which $\ba^z = 0$ satisfies our
requirements; if $\ba^2 \geq 0$, the required frame is one in which
$A^t$ is non-zero but sufficiently small that $\fc_1^2 > \fc_1 \fc_4
(\ba^t)^2$.)  For $\ba^2 \neq 0$, this is equivalent to the condition
that the ``effective metric'' appearing in \eqref{Maxwellian} is of
signature $(- \: + \: + \: + )$ or $(+ \: - \: - \: - )$.

\begin{widetext}
\section{Tensor coefficient-matching}
\label{nastytensors}

For a general vector theory, we will have
\begin{multline}
  \label{Q0full}
  (\mathcal{Q}_{K0}) {}^{abc} {}_d {}^e = (\mc_1 - \mc_3) \ba^{(a}
  \eta^{b)c} \delta^e {}_d + (\mc_3 - \mc_1) \ba^{(a} \delta^{b)} {}_d
  \eta^{ce} - (\mc_1 + \mc_3 ) \delta^{(a} {}_d \eta^{b)c} \ba^e \\
  + \left( \mc_4 - \frac{1}{2} \mc_5 \right) \ba^a \ba^b \ba^c
  \delta^e {}_d - 2 \mc_4 \ba^{(a} \delta^{b)} {}_d \ba^c \ba^e +
  \left( \frac{1}{2} \mc_5 - \mc_6 \right) \ba^a \ba^b \eta^{ce} \ba_d
  \\ 
  - \mc_5 \ba^{(a} \eta^{b)c} \ba_d \ba^e - \frac{1}{2} \mc_7 \ba^a
  \ba^b \delta^c {}_d \ba^e - \frac{1}{2} \mc_7 \eta^{ab} \ba^c \ba_d
  \ba^e - \mc_8 \ba^a \ba^b \ba^c \ba_d \ba^e.
\end{multline} 
Assuming that $K^{(ab)c} {}_d \neq 0$, the tensor
$\mathcal{T}_0^{abc}$ must have $U_1 = 0$ and $U_2 + U_3 A^2 + 1 = 0$;
multiplying these two tensors together, we find that
\begin{multline}
  \label{TK0full}
  \mathcal{T}_0^{abf} (K_0)^e {}_f {}^c {}_d = U_2 \mc_3 \ba^{(a}
  \eta^{b)c} 
  \delta^e {}_d + U_2 \mc_1 \ba^{(a} \delta^{b)} {}_d \eta^{ce} -
  \left( \frac{1}{2} \mc_5 + \ba^{-2} (1 + U_2) \mc_3 \right) \ba^a
  \ba^b \ba^c \delta^e {}_d \\ - \frac{1}{2} \mc_7 \ba^a \ba^b
  \delta^c {}_d \ba^e + U_2 \mc_4 \ba^{(a} \delta^{b)} {}_d \ba^c
  \ba^e - \left( \mc_6 + \ba^{-2} (1 + U_2) \mc_1 \right) \ba^a \ba^b
  \eta^{ce} \ba_d 
  + \frac{1}{2} U_2 \mc_5 \ba^{(a} \eta^{b)c} \ba_d \ba^e \\ +
  \frac{1}{2} U_2 
  \mc_7 \ba^{(a} \eta^{b)e} \ba^c \ba_d
  - \left( \mc_8 + \ba^{-2} (1 + U_2) \left( \mc_4 + \frac{1}{2}
      (\mc_5 + \mc_7) \right) \right) \ba^a \ba^b \ba^c \ba_d \ba^e.
\end{multline}

For a pseudo-Maxwell vector theory, we can obtain
$(\mathcal{Q}_{K0})^{abc} {}_d {}^e$ simply by applying the conditions
\eqref{maxfns} to \eqref{Q0full}; the result is
\begin{equation}
  \label{Q0maxfull}
  (\mathcal{Q}_{K0}) {}^{abc} {}_d {}^e = 
  2 \mc_1 \ba^{(a} \eta^{b)c} \delta^e {}_d 
  - 2 \mc_1 \ba^{(a} \delta^{b)} {}_d \eta^{ce} 
  + 2 \mc_4 \ba^a \ba^b \ba^c \delta^e {}_d 
  - 2 \mc_4 \ba^{(a} \delta^{b)} {}_d \ba^c \ba^e \\
  - 2 \mc_4 \ba^a \ba^b \eta^{ce} \ba_d
  + 2 \mc_4 \ba^{(a} \eta^{b)c} \ba_d \ba^e. 
\end{equation}
Due to the massive-mode condition, however, the above constraints on
the functions $U_i$ are relaxed;  we thus must allow for arbitrary
$U_i$ functions, yielding
\begin{multline}
  \label{TK0maxfull}
  \mathcal{T}_0^{abf} (K_0)^e {}_f {}^c {}_d = U_2 \mc_1( \ba^{(a}
  \delta^{b)} {}_d \eta^{ce} - \ba^{(a} \eta^{b)c} \delta^e {}_d) +
  U_1 (\mc_1 + \ba^2 \mc_4) \eta^{ab} (\eta^{ce} \ba_d - A^c \delta^e
  {}_d) \\ + \left( U_2 \mc_4 + U_3 (\mc_1 + \ba^2 \mc_4) \right) \ba^a
  \ba^b
  ( \eta^{ce} \ba_d - \ba^c \delta^e {}_d)
  + U_2 \mc_4 (\ba^{(a} \delta^{b)} {}_d \ba^c \ba^e - \ba^{(a}
  \eta^{b)c} \ba_d \ba^e).
\end{multline}

At $\mathcal{O}(\xi)$, we can attempt an analogous coefficient
matching for the tensors in Equation \eqref{Qcondxi}.  The
left-hand side of \eqref{Qcondxi} is given by
\begin{multline}
  \label{tensmatch1}
  \mathcal{Q}_R {}^{ab(c} {}_d {}^{e)} + \hat{\mathcal{Q}}_K
  {}^{ab(c} {}_d {}^{e)}
  = (\md_1 - \md_3 + f_s) \ba^{(a} \eta^{b)(c} \delta^{e)} {}_d 
  + (\md_3 - \md_1 - f_s) \ba^{(a} \delta^{b)} {}_d \eta^{ce}
  + (-\md_1 - \md_3 + f_s) \delta^{(a} {}_d \eta^{b)(c} \ba^{e)} \\
  + 2 f'_u \eta^{ab} \eta^{ce} \ba_d 
  - 2 f'_u \eta^{e(a} \eta^{b)c} \ba_d 
  - f_s \eta^{ab} \ba^{(c} \delta^{e)} {}_d 
  + \left( \md_4 - \frac{1}{2}(\md_5 + \md_7) \right )\ba^a \ba^b
  \ba^{(c} \delta^{e)} {}_d 
  - 2 \md_4 A^{(a} \delta^{b)} {}_d \ba^c \ba^e \\
  + \left( \frac{1}{2} \md_5 - \md_6 - f'_s \right) \ba^a \ba^b
  \eta^{ce} \ba_d 
  - (\md_5 - 2f'_s) \ba^{(a} \eta^{b)(c} \ba^{e)} \ba_d
  - \left( \frac{1}{2} \md_7 + f'_s \right) \eta^{ab} \ba^c \ba^e
  \ba_d 
  - \md_8 \ba^a \ba^b \ba^c \ba^e \ba_d,
\end{multline}
and the right-hand side is given by
\begin{multline}
  \label{tensmatch2}
  (V_2 \mc_3 - 2 \md_3) \ba^{(a} \eta^{b)(c} \delta^{e)} {}_d 
  + (V_2 \mc_1\ - 2 \md_1) \ba^{(a} \delta^{b)} {}_d \eta^{ce}
  + V_1 (\mc_1 + \ba^2 \mc_6) \eta^{ab} \eta^{ce} \ba_d \\
  + \left( \frac{1}{2} V_2 \mc_5 - (\md_5 + \md_7) + V_3
    \left( \mc_3 + \frac{1}{2} \ba^2 \mc_5  \right) + U_3
    \left( \md_3 + \frac{1}{2} \ba^2 (\md_5 + \md_7) \right) \right)
  \ba^a \ba^b \ba^{(c} \delta^{e)} {}_d \\
  + V_1 \left( \mc_3 + \frac{1}{2} \ba^2 \mc_5  \right)
  \eta^{ab} \ba^{(c} \delta^{e)} {}_d 
  + (V_2 \mc_4 - 2 \md_4) A^{(a} \delta^{b)} {}_d \ba^c \ba^e 
  + \left( \frac{1}{2} V_2 \mc_5  - \md_5 - \md_7 \right) \ba^{(a}
  \eta^{b)(c} \ba^{e)} \ba_d  \\
  + (V_2 \mc_6 - 2 \md_6 + V_3 (\mc_1 + \ba^2 \mc_6) + U_3 (\md_1 +
  \ba^2 \md_6) ) \ba^a \ba^b \eta^{ce} \ba_d
  + V_1 \left( \mc_4 + \frac{1}{2} \mc_5 + \ba^2 \mc_8 \right)
  \eta^{ab} \ba^c \ba^e \ba_d  \\ 
  + \left( V_2 \mc_8 - 2 \md_8 + V_3 \left( \mc_4 + \frac{1}{2} \mc_5
      + \ba^2 \mc_8 \right) + U_3 \left( \md_4 + \frac{1}{2} (\md_5 +
      \md_7) + \ba^2 \md_8 \right) \right) \ba^a \ba^b \ba^c \ba^e
  \ba_d.
\end{multline}
We have used the fact that both candidate vector kinetic terms
found in the previous section have $U_1 = 0$, $U_2 = -2$ and
$\mc_7 = 0$.  
\end{widetext}

\bibliography{vectorLV}{}

\end{document}